\RequirePackage{fix-cm}
\documentclass[twocolumn]{svjour3}          
\smartqed  
\usepackage{graphicx}
\usepackage{mathptmx}      
\usepackage{natbib}
\usepackage{color}
\journalname{Microgravity Science and Technology}

\usepackage{amsmath}
\usepackage{amssymb}

\begin{document}

\title{Comparison of the effect of horizontal vibrations on interfacial waves in a two-layer system of inviscid liquids to effective gravity inversion}
\titlerunning{Horizontal vibrations and effective gravity inversion for two-layer systems of inviscid liquids}
\author{Anastasiya V.\ Pimenova$^{1}$ \and Denis S.\ Goldobin$^{1,2}$ \and Tatyana P.\ Lyubimova$^{1,2}$}
\authorrunning{A.V.\ Pimenova, D.S.\ Goldobin, T.P.\ Lyubimova}
\institute{$^1$Institute of Continuous Media Mechanics, UB RAS,
         1 Academik Korolev str., Perm 614013, Russia\\
         $^2$Department of Theoretical Physics,
         Perm State University, 15 Bukireva str., 614990, Perm, Russia}
\date{Received: date / Accepted: date}

\maketitle

\begin{abstract}
We study the waves at the interface between two thin horizontal layers of immiscible liquids subject to high-frequency tangential vibrations. Nonlinear governing equations are derived for the cases of two- and three-dimensional flows and arbitrary ratio of layer thicknesses. The derivation is performed within the framework of the long-wavelength approximation, which is relevant as the linear instability of a thin-layers system is long-wavelength. The dynamics of equations is integrable and the equations themselves can be compared to the Boussinesq equation for the gravity waves in shallow water, which allows one to compare the action of the vibrational field to the action of the gravity and its possible effective inversion.
\keywords{Interfacial waves \and Two-layer liquid system \and Longitudinal vibrations \and Boussinesq equation}
\end{abstract}

\renewcommand{\Re}{\mathrm{Re}}
\renewcommand{\Im}{\mathrm{Im}}

\section{Introduction}
\label{sec_intro}
The first experimental studies on the usage of vibrations for stabilizing otherwise unstable configurations of multiphase fluid systems were reported by~\citet{Wolf-1961, Wolf-1970}. Under the weightlessness conditions, the demand for such a control tool even increases because of the necessity to maintain stratification of fluids or control convection for diverse technological systems, which is well highlighted by the ongoing research on the subject~\citep{Thiele-Vega-Knobloch-2006, Mialdun-etal-2008, Shklyaev-Alabuzhev-Khenner-2009, Nepomnyashchy-Simanovskii-2013, Gaponenko-Shevtsova-2016, Bratsun-etal-2016, Lappa-2016, Smorodin-Myznikova-Keller-2017, Lyubimova-etal-2017}.

\begin{figure}[t]
\centerline{
 \includegraphics[width=0.46\textwidth]%
 {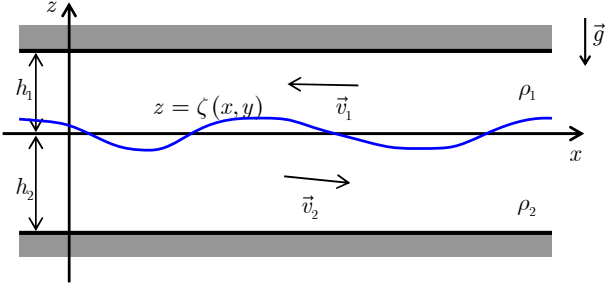}
}

  \caption{
Sketch of the coordinate frame and two-layer liquid system subject to longitudinal vibrations.
 }
  \label{fig1}
\end{figure}

Wolf's experimental observations of wave patterns on the interface between immiscible fluids subject to horizontal vibrations received their first solid theoretical basis with the linear instability analysis of the flat state of the interface~\citep{Lyubimov-Cherepanov-1986, Khenner-Lyubimov-Shotz-1998, Khenner-etal-1999}.
In Fig.~\ref{fig1}, one can see the sketch of the system for which the instability was theoretically revealed for strong enough vibrations. La\-ter on, the nonlinear dynamics of interfacial waves for the high-viscosity case and one liquid layer was studied analytically by \citet{Shklyaev-Alabuzhev-Khenner-2009} and \citet{Benilov-Chugunova-2010}. In Wolf's experiments~\citep{Wolf-1961,Wolf-1970}, the viscous boundary layer in the most viscous liquid was an order of magnitude thinner than the liquid layer, meaning the approximation of inviscid liquid to be relevant. The analytical treatment of the nonlinear dynamics of interfacial waves in inviscid liquids was made possible in \citep{Goldobin-etal-EPL-2014-solitons, Goldobin-etal-PRE-2015}, where the governing equations for two-dimensional flows were reported. The equations were derived below the instability threshold within the framework of the long-wa\-ve\-length approximation for the case of equal thickness of layers. Simultaneously, the understanding of strongly nonlinear regimes of the dynamics of the interface in {\it low-viscosity} liquids above the instability threshold was significantly advanced in~\citep{Lyubimova-etal-2017} by means of numerical simulation accompanied by analytical estimates.

In this paper we extend the derivation of governing equations for inviscid liquids to the case of arbitrary ratio of layer thicknesses and admit the variation of flow in the horizontal direction orthogonal to vibrations, i.e., consider a 3-d problem. The equations to be derived for both `normal' and `inverted' (the heavy liquid overlies the light one) configurations of the system. The comparison of these equations with the Boussinesq equation for the gravity waves in shallow water is of interest and suggests noteworthy interpretations of the action of the vibration field on the system.

\section{Problem statement and governing equations}
\label{sec_statement}
We consider a system of two horizontal layers of immiscible inviscid liquids, confined between two impermeable horizontal boundaries (see Fig.~\ref{fig1}). The system is subject to high-frequency longitudinal vibrations of linear polarization; the velocity of vibrational motion of the system is $(b/2)e^{i\omega t}+c.c.$ (here and hereafter, ``$c.c.$'' stands for complex conjugate). The density of the upper liquid $\rho_1$ is smaller than the density of the lower one $\rho_2$. The layer thicknesses are $h_1$ and $h_2$ (see Fig.~\ref{fig1}). We choose the horizontal coordinate $x$ along the direction of vibrations, the $z$-axis is vertical with origin at the unperturbed interface between layers.

In this system, at the limit of infinitely long layers, the state with flat interface $z=\zeta(x,y)=0$ is always possible. If the layers are limited in horizontal directions by impermeable lateral boundaries, the interface will be nearly flat at a distance from these boundaries. For inviscid fluids, this state (the ground state) is featured by spatially homogeneous pulsating velocity fields $\vec{v}_{j0}$ in both layers;
\begin{equation}
\begin{array}{l}
\displaystyle
\vec{v}_{j0}=a_j(t)\vec{e}_x,\qquad
 a_j(t)=A_je^{i\omega t}+c.c.,
\\[10pt]
\displaystyle
 A_1=\frac{(h_1+h_2)\rho_2 b}{2(h_2\rho_1+h_1\rho_2)},\quad
 A_2=\frac{(h_1+h_2)\rho_1 b}{2(h_2\rho_1+h_1\rho_2)},
\end{array}
\label{eq01}
\end{equation}
where $j=1,2$ and $\vec{e}_x$ is the unit vector of the $x$-axis. (The shift of the time offset results in a complex multiplier for $b$ and $A_j$; therefore, one can choose the time offset so that $b$ and $A_j$ will be real.) The result (\ref{eq01}) follows from the condition of zero pressure jump across the uninflected interface and the condition of the total fluid flux through the vertical cross-section which is
$\int_{-h_2}^{+h_1}v^{(x)}dz=(h_1+h_2)b\cos{\omega t}$
(which is due to the system motion with velocity $b\cos{\omega t}$).

Considering flow of inviscid liquid, it is convenient to introduce potential $\phi_j$ of the velocity field;
\begin{equation}
\vec{v}_j=-\nabla\phi_j\,.
\label{eq02}
\end{equation}
The mass conservation law for incompressible liquid, $\nabla\cdot\vec{v}_j=0$, yields the Laplace equation for potential, $\Delta\phi_j=0$. The kinematic conditions on the top and bottom boundaries
\begin{equation}
\phi_{1z}(z=h_1)=\phi_{2z}(z=-h_2)=0
\label{eq03}
\end{equation}
and on the interface $z=\zeta(x,y)$
\begin{eqnarray}
\dot{\zeta}&=&-\phi_{1z}+\nabla\phi_1\cdot\nabla\zeta\,,
\label{eq04}
\\[10pt]
\dot{\zeta}&=&-\phi_{2z}+\nabla\phi_2\cdot\nabla\zeta
\label{eq05}
\end{eqnarray}
are also to be taken into account. (Here and hereafter, the upper dot stands for the time-derivative and letter in subscript denotes partial derivative with respect to the corresponding coordinate.) Eqs.~(\ref{eq04}) and (\ref{eq05}) can be derived from the condition that the points of zero value of the distance function $F=z-\zeta(x,y)$, which correspond to the position of the interface, move with liquid, i.e., the Lagrangian derivative (material derivative)
 $dF/dt=\partial F/\partial t+\vec{v}\cdot\nabla{F}$
is zero on the interface: $-\dot\zeta+v^{(z)}-\vec{v}\cdot\nabla\zeta=0$, and this holds for both liquids.

After substitution of the potential flow, the Euler equation takes the following form:
\[
\nabla\left(-\dot{\phi}_j+\frac{1}{2}\left(\nabla\phi_j\right)^2\right)
 =\nabla\left(-\frac{1}{\rho_j}p_j-gz\right),
\]
where $g$ is the gravity acceleration. The latter equation provides the expression for the pressure field in the volume of two liquids for a given flow field;
\begin{equation}
p_j=p_{j0}+\rho_j\left(\dot{\phi}_j
 -\frac{1}{2}\left(\nabla\phi_j\right)^2-gz\right).
\label{eq06}
\end{equation}

Now the stress on the interface is remaining to be brought into account to make the equation system self-contained by providing required boundary conditions for $\phi_j$ on the interface between two liquids. The pressure jump across the interface is caused by the surface tension;
\begin{equation}
z=\zeta(x,y):
\;
p_1-p_2 =-\alpha\nabla\cdot\vec{n}\;
 \left(\mbox{where  }\vec{n}:=\frac{\nabla F}{\left|\nabla F\right|}\right),
\label{eq07}
\end{equation}
where $\alpha$ is the surface tension coefficient and $\vec{n}$ is
the unit vector normal to the interface.

The linear stability analysis revealed the marginal vi\-bra\-tion-induced instability of the flat-interface state to be long-wa\-ve\-length~\citep{Lyubimov-Cherepanov-1986,Goldobin-etal-PRE-2015}. Hence, we restrict our consideration to the case of the long-wa\-ve\-length approximation, $\left|{\partial_x\vec{v}}\right|\ll\left|{\partial_z\vec{v}}\right|$.

\section{Governing equations for long-wavelength patterns}\label{sec_deriv}
\subsection{Derivation of equations for 2-dimensional flow}
In this section we derive the governing equation for long-wa\-ve\-length patterns. We employ the standard method of multiple scales with small parameters $\omega^{-1}$ and $l^{-1}$, where $l$ is the reference horizontal length of patterns, $\partial_x\sim l^{-1}$. The hierarchy of small parameters and the orders of magnitude of fields will be established in the course of derivation.

Within the long-wavelength approximation, the solutions to the Laplace equation for $\phi_j(x,t)$ satisfying boundary conditions~(\ref{eq03}) in the most general form read
\begin{equation}
\begin{array}{r}
\displaystyle
\phi_1=-a_1(t)x+\Phi_1(x,t)-\frac{1}{2}(h_1-z)^2\Phi_{1xx}(x,t)
\qquad\\[10pt]
\displaystyle
 {}+\frac{1}{4!}(h_1-z)^4\Phi_{1xxxx}(x,t)-\dots\,,
\end{array}
\label{eqa01}
\end{equation}
\begin{equation}
\begin{array}{r}
\displaystyle
\phi_2=-a_2(t)x+\Phi_2(x,t)-\frac{1}{2}(h_2+z)^2\Phi_{2xx}(x,t)
\qquad\\[10pt]
\displaystyle
 {} +\frac{1}{4!}(h_2+z)^4\Phi_{2xxxx}(x,t)-\dots\,.
\end{array}
\label{eqa02}
\end{equation}
Here the ground state (the flat-interface state) is represented by
the terms $-a_j(t)x$; $\Phi_j(x,t)$ describe perturbation flow,
they are yet arbitrary functions of $x$ and $t$. After
substitution of $p_j$ from expression~(\ref{eq06}) and $\phi_j$
from expressions~(\ref{eqa01})--(\ref{eqa02}), the condition of
stress balance on the interface (\ref{eq07}) reads
\begin{eqnarray}
&&
 p_{1\infty}-p_{2\infty}
 +\rho_1\Bigg[-\dot{a}_1x+\dot{\Phi}_1-\frac{(h_1-\zeta)^2}{2}\dot{\Phi}_{1xx}
\nonumber\\[5pt]
&&
 \qquad\qquad
 {}-\frac{1}{2}\left(-a_1+\Phi_{1x}-\frac{(h_1-\zeta)^2}{2}\Phi_{1xxx}\right)^2
\nonumber\\[5pt]
&&
 \qquad\qquad\qquad
 {}-\frac{((h_1-\zeta)\Phi_{1xx})^2}{2}+\dots\Bigg]
\nonumber\\[5pt]
&&
 \qquad
 {}-\rho_2\Bigg[-\dot{a}_2x+\dot{\Phi}_2-\frac{(h_2+\zeta)^2}{2}\dot{\Phi}_{2xx}
\nonumber\\[5pt]
&&
 \qquad\qquad
 {}-\frac{1}{2}\left(-a_2+\Phi_{2x}-\frac{(h_2+\zeta)^2}{2}\Phi_{2xxx}\right)^2
\nonumber\\[5pt]
&&
 \qquad\qquad\qquad
 {}-\frac{((h_2+\zeta)\Phi_{2xx})^2}{2}+\dots\Bigg]
\nonumber\\[5pt]
&&
 \qquad
 {}+(\rho_2-\rho_1)g\zeta=\alpha\frac{\zeta_{xx}}{(1+\zeta_x^2)^{3/2}}\,.
\nonumber
\end{eqnarray}
Here ``\dots'' stand for terms
$\mathcal{O}_1(\dot\Phi_jh_j^4/l^4)+\mathcal{O}_2(a_j\Phi_jh_j^4/l^5)+\mathcal{O}_3(\Phi_j^2h_j^4/l^6)$.
The difference of constants $p_{1\infty}-p_{2\infty}$ is to be determined from the condition that in the area of vanishing perturbations of the pulsation flow, i.e.\ $\Phi_j(x,t)=const$, the interface remains flat, i.e.\ $\zeta(x,t)=0$. This condition yields
$p_{1\infty}-p_{2\infty}-(\rho_1a_1^2(t)-\rho_2a_2^2(t))/2=0$.
Choosing measure units for length: $L=\sqrt{\alpha/[(\rho_2-\rho_1)g]}$, for time: $T=L/b$, and for the fluid densities: $\rho_\ast$---which means replacement
\begin{equation}
\begin{array}{c}
\displaystyle
(x,z)\to(Lx,Lz)\,,\qquad t\to Tt\,,\qquad \zeta\to L\zeta\,,
\\[5pt]
\displaystyle
\Phi_j\to(L^2/T)\Phi_j\,,\qquad \rho_i\to\rho_\ast\rho_i
\end{array}
\label{rescaling1}
\end{equation}
in equations---one can rewrite the last equation in the dimensionless form
\begin{gather}
 B\Bigg[\frac{\rho_1a_1^2-\rho_2a_2^2}{2}+\rho_1\dot{\Phi}_1
 -\frac{\rho_1(h_1-\zeta)^2}{2}\dot{\Phi}_{1xx}
 -\frac{\rho_1}{2}\bigg(a_1
\nonumber
 \\[8pt]
 {}
 -\Phi_{1x}+\frac{1}{2}(h_1-\zeta)^2\Phi_{1xxx}\bigg)^2
 -\frac{\rho_1}{2}\left((h_1-\zeta)\Phi_{1xx}\right)^2
 -\rho_2\dot{\Phi}_2
\nonumber
 \\[8pt]
 {}
 +\frac{\rho_2(h_2+\zeta)^2}{2}\dot{\Phi}_{2xx}
 +\frac{\rho_2}{2}\bigg(a_2-\Phi_{2x}+\frac{1}{2}(h_2+\zeta)^2\Phi_{2xxx}\bigg)^2
\nonumber
 \\[8pt]
 {}
 +\frac{\rho_2}{2}\left((h_2+\zeta)\Phi_{2xx}\right)^2
 +\dots\Bigg]+\zeta=\frac{\zeta_{xx}}{(1+\zeta_x^2)^{3/2}}\,.
\label{eqa03}
\end{gather}
Here the dimensionless vibration parameter
\begin{equation}
 B\equiv\frac{\rho_\ast b^2}{\sqrt{\alpha(\rho_2-\rho_1)g}}=B_0+B_1
\label{eqa04}
\end{equation}
($\rho_j$ is dimensional here), where $B_0 $ is the critical value of the vibration parameter above which the flat-interface state becomes linearly unstable, $B_1$ is a small deviation of the vibration parameter from the critical value. Further, kinematic conditions~(\ref{eq04}) and (\ref{eq05}) turn into
\begin{equation}
\dot\zeta=\left(-(h_1-\zeta)\Phi_{1x}+\frac{1}{3!}h_1^3\Phi_{1xxx}
 -a_1\zeta+\dots\right)_x ,
\label{eqa05}
\end{equation}
\begin{equation}
\dot\zeta=\left((h_2+\zeta)\Phi_{2x}-\frac{1}{3!}h_2^3\Phi_{2xxx}
 -a_2\zeta+\dots\right)_x .
\label{eqa06}
\end{equation}
Here ``\dots'' stand for
$\mathcal{O}_1(\Phi_jh_j^2\zeta/l^3)+\mathcal{O}_2(\Phi_jh_j^4/l^5)$.
Eqs.~(\ref{eqa03}), (\ref{eqa05}), and (\ref{eqa06}) form a self-contained equation system.

It is convenient to distinguish two main time-modes in fields: the average over vibration period part and the pulsation part;
\[
\begin{array}{l}
 \zeta=\eta(\tau,x)+\xi(\tau,x)e^{i\omega t}+c.c.+\dots\,, \\[10pt]
 \Phi_j=\varphi_j(\tau,x)+\psi_j(\tau,x)e^{i\omega t}+c.c.+\dots\,,
\end{array}
\]
where $\tau$ is a ``slow'' time related to the average over vibration period evolution and ``\dots'' stand for higher powers of $e^{i\omega t}$.

In order to develop an expansion in small parameter $\omega^{-1}$, we have to adopt certain hierarchy of smallness of parameters, fields, etc. We adopt small deviation from the instability threshold $B_1\sim\omega^{-1}$. Then $\eta\sim\omega^{-1}$ and $\partial_x\sim\omega^{-1/2}$~\citep{Lyubimov-Cherepanov-1986,Goldobin-etal-PRE-2015}. It is as well established  \citep{Lyubimov-Cherepanov-1986} that for finite wavelength perturbations (finite $k\ne0$) $B_0(k)=B_0(0)+Ck^2+\mathcal{O}(k^4)$. Generally, the expansion of exponential growth rate of perturbations in series of $B_1$ near the instability threshold possesses a non-zero linear part, and $B_0(k)-B_0(0)\sim k^2$; therefore, $\partial_\tau\sim\mathcal{O}_1(B_1)+\mathcal{O}_2(k^2)\sim\omega^{-1}$. The order of magnitude of $\xi$, $\varphi_j$ and $\psi_j$ is more convenient to be determined in the course of development of the expansion.

Collecting in Eqs.~(\ref{eqa05})--(\ref{eqa06}) terms with $e^{i\omega t}$, one finds
\begin{gather}
i\omega\xi+\xi_\tau=\Big(-(h_1-\eta)\psi_{1x}
\nonumber\\[5pt]
\qquad\qquad\qquad{}
 +\frac{1}{3!}h_1^3\psi_{1xxx} +\xi\varphi_{1x}-A_1\eta+\dots\Big)_x\,,
\label{eqa07}
\\[5pt]
i\omega\xi+\xi_\tau=\Big((h_2+\eta)\psi_{2x}
\nonumber\\[5pt]
\qquad\qquad\qquad{}
 -\frac{1}{3!}h_2^3\psi_{2xxx}
 +\xi\varphi_{2x}-A_2\eta+\dots\Big)_x\,,
\label{eqa08}
\end{gather}
where ``\dots'' stand for
$\mathcal{O}_1((\xi\varphi+\eta\psi)h_j^2/l^4)+\mathcal{O}_2(\psi\,h_j^4/l^6)$.
Constant with respect to $t$ terms sum-up to
\begin{equation}
\eta_\tau=\Big(-(h_1-\eta)\varphi_{1x}+\xi\psi_{1x}^\ast+c.c.-A_1\xi^\ast+c.c.+\dots\Big)_x\,,
\label{eqa09}
\end{equation}
\begin{equation}
\eta_\tau=\Big((h_2+\eta)\varphi_{2x}+\xi\psi_{2x}^\ast+c.c.-A_2\xi^\ast+c.c.+\dots\Big)_x\,,
\label{eqa10}
\end{equation}
where the superscript ``$*$'' stands for complex conjugate and ``\dots'' stand for $\mathcal{O}_1((\eta\varphi+\xi\psi)h_j^2/l^4)+\mathcal{O}_2(\varphi h_j^4/l^6)$. The difference of (\ref{eqa07}) and (\ref{eqa08}) yields $\psi_j\sim\omega^{-1/2}$, the difference of (\ref{eqa09}) and (\ref{eqa10}) yields $\varphi_j\sim\omega^{-1}$. For dealing with non-linear terms in the consideration that follows, it is convenient to extract the first correction to $\psi_j$ explicitly, i.e.\ write
$\psi_j=\psi_j^{(0)}+\psi_j^{(1)}+\dots$\,, where
$\psi_j^{(1)}\sim\omega^{-1}\psi _j^{(0)}\sim\omega^{-3/2}$.
Eq.~(\ref{eqa07}) (or (\ref{eqa08})) yields in the leading order ($\sim\omega^{-3/2}$)
\begin{equation}
\xi=\frac{i}{\omega}(h_1\psi_{1x}+A_1\eta)_x=
 -\frac{i}{\omega}(h_2\psi_{2x}+A_2\eta)_x\sim\omega^{-\frac{5}{2}}\,.
\label{eqa11}
\end{equation}

Considering the difference of (\ref{eqa08}) and (\ref{eqa07}), one has to keep in mind, that we are interested in localized patterns for which $\Phi_{jx}(x=\pm\infty)=0$, $\zeta(x=\pm\infty)=0$. Hence, this difference can be integrated with respect to $x$, taking the form
$$
\begin{array}{r}
\displaystyle
h_1\psi_{1x}+h_2\psi_{2x}-\eta(\psi_1-\psi_2)_x
 -\frac{1}{6}(h_1^3\psi_1+h_2^3\psi_2)_{xxx}
\qquad\\[10pt]
\displaystyle
 {}-\xi(\varphi_1-\varphi_2)_x+(A_1-A_2)\eta+\dots=0\,,
\end{array}
$$
which yields in the first two orders of smallness
\begin{eqnarray}
(h_1\psi_1^{(0)}+h_2\psi_2^{(0)})_x=-(A_1-A_2)\eta
\qquad\qquad\qquad\nonumber\\[5pt]
=-\frac{(\rho_2-\rho_1)(h_1+h_2)}{2(h_1\rho_2+h_2\rho_1)}\eta\,,
\label{eqa12}
\end{eqnarray}
\begin{eqnarray}
(h_1\psi_1^{(1)}+h_2\psi_2^{(1)})_x=(\psi_1^{(0)}-\psi_2^{(0)})_x\eta
\qquad\qquad\quad\nonumber\\[5pt]
{}+\frac{1}{6}(h_1^3\psi_1^{(0)}+h_2^3\psi_2^{(0)})_{xxx}\,.
\label{eqa13}
\end{eqnarray}
The difference and the sum of Eqs.~(\ref{eqa09}) and (\ref{eqa10}) yield in the leading order, respectively,
\begin{equation}
h_1\varphi_1=-h_2\varphi_2\,,
\label{eqa14}
\end{equation}
\begin{equation}
\eta_\tau=-h_1\varphi_{1xx}=h_2\varphi_{2xx}\,.
\label{eqa15}
\end{equation}

Let us now consider Eq.~(\ref{eqa03}). We will collect groups of terms with respect to power of $e^{i\omega t}$ and the order of smallness in $\omega^{-1}$.
\\
$\underline{\sim\omega^{+\frac{1}{2}}e^{i\omega t}}$:
$$
i\omega B_0(\rho_1\psi_1^{(0)}-\rho_2\psi_2^{(0)})=0\,.
$$
We introduce
\begin{equation}
\psi^{(0)}\equiv\rho_j\psi_j^{(0)}\,.
\label{eqa16}
\end{equation}
The last equation and equation~(\ref{eqa12}) yield
\begin{equation}
\psi_x^{(0)}=-\frac{\rho_1\rho_2(\rho_2-\rho_1)(h_1+h_2)}{2(h_1\rho_2+h_2\rho_1)^2}\eta\,.
\label{eqa17}
\end{equation}
The next group of terms to be formally collected should be formed by the contributions of the order $\omega^0$ and proportional to $e^{i\omega t}$, i.e.,
\\
$\underline{\sim\omega^0e^{i\omega t}}$:
$$
\mbox{No contributions.}
$$
Further,
\\
$\underline{\sim\omega^{-\frac{1}{2}}e^{i\omega t}}$:
\[
\begin{array}{l}
\displaystyle
 i\omega B_1\underbrace{(\rho_1\psi_1^{(0)} -\rho_2\psi_2^{(0)})}_{\quad=0}
 +i\omega B_0(\rho_1\psi_1^{(1)}-\rho_2\psi_2^{(1)})
\\
\displaystyle
 {}+B_0\underbrace{(\rho_1\psi_1^{(0)}-\rho_2\psi_2^{(0)})_\tau}_{\quad=0}
\\
\displaystyle
\quad
 {}+i\omega B_0\frac{1}{2}(h_2^2\rho_2\psi_{2xx}^{(0)} -h_1^2\rho_1\psi_{1xx}^{(0)})=0\,.
 \end{array}
\]
(We marked the combinations which are known to be zero from the leading order of expansion.) Hence,
\begin{equation}
\rho_1\psi_1^{(1)}-\rho_2\psi_2^{(1)}=\frac{1}{2}(h_1^2-h_2^2)\psi_{xx}^{(0)}.
\label{eqa18}
\end{equation}
The last equation and equation~(\ref{eqa13}) yield
\begin{eqnarray}
\psi_{1x}^{(1)}&=&\frac{\rho_2-\rho_1}{\rho_1(h_1\rho_2+h_2\rho_1)}\psi_x^{(0)}\eta
 \nonumber\\[5pt]
 &&\qquad{}
 +\frac{3h_2h_1^2\rho_1-2h_2^3\rho_1+h_1^3\rho_2}{6\rho_1(h_1\rho_2+h_2\rho_1)}\psi_{xxx}^{(0)}
 \nonumber\\[5pt]
 &=&-\frac{\rho_2(\rho_2-\rho_1)^2(h_1+h_2)}{2(h_1\rho_2+h_2\rho_1)^3}\eta^2
\nonumber\\[5pt]
 &&\hspace{-30pt}
 {}-\frac{\rho_2(3h_2h_1^2\rho_1-2h_2^3\rho_1+h_1^3\rho_2)(\rho_2-\rho_1)(h_1+h_2)}
 {12\,(h_1\rho_2+h_2\rho_1)^3}\eta_{xx}\,,
\label{eqa19}
\end{eqnarray}
\begin{eqnarray}
\psi_{2x}^{(1)}&=&-\frac{\rho_1(\rho_2-\rho_1)^2(h_1+h_2)}{2(h_1\rho_2+h_2\rho_1)^3}\eta^2
\nonumber\\[5pt]
 &&\hspace{-30pt}
 {}-\frac{\rho_1(3h_1h_2^2\rho_2-2h_1^3\rho_2+h_2^3\rho_1)(\rho_2-\rho_1)(h_1+h_2)}
 {12\,(h_1\rho_2+h_2\rho_1)^3}\eta_{xx}\,.
\label{eqa19-2}
\end{eqnarray}
$\underline{\sim\omega^{-1}(e^{i\omega t})^0}$:
\begin{equation}
B_0[-\rho_2(A_2\psi_{2x}^{(0)\ast}+c.c.)+\rho_1(A_1\psi_{1x}^{(0)\ast}+c.c.)]+\eta=0\,.
\label{eqa20-0}
\end{equation}
Substituting~(\ref{eqa16}) and (\ref{eqa17}) into the last equation, one finds
$$
\left[-\frac{B_0\rho_1\rho_2(\rho_2-\rho_1)^2(h_1+h_2)^2}{2(h_1\rho_2+h_2\rho_1)^3}+1\right]\eta=0\,.
$$
Thus we obtain the solvability condition, which poses a restriction on $B_0$; this restriction determines the linear instability threshold
\begin{equation}
B_0=\frac{2(h_1\rho_2+h_2\rho_1)^3}{\rho_1\rho_2(\rho_2-\rho_1)^2(h_1+h_2)^2}\,.
\label{eqa20}
\end{equation}
Since the threshold $B_0$ is an important and experimentally measurable characteristic of the system, we provide it here also in original {\it dimensional} terms (cf.\ Eqs.~(\ref{rescaling1}) and (\ref{eqa04})):
\begin{equation}
 B_0=\frac{2\rho_\ast(h_1\rho_2+h_2\rho_1)^3}{\rho_1\rho_2(\rho_2-\rho_1)^2(h_1+h_2)^2}
 \sqrt{\frac{(\rho_2-\rho_1)g}{\alpha}}\,,
\nonumber
\end{equation}
\begin{equation}
 b_0^2=\frac{2(h_1\rho_2+h_2\rho_1)^3g}{\rho_1\rho_2(\rho_2-\rho_1)(h_1+h_2)^2}\,.
\nonumber
\end{equation}

$\underline{\sim\omega^{-2}(e^{i\omega t})^0}$:
\begin{equation}
\begin{array}{l}
\displaystyle
 B_1\underbrace{[-\rho_2(A_2\psi_{2x}^{(0)\ast}+c.c.)
                +\rho_1(A_1\psi_{1x}^{(0)\ast}+c.c.)]}_{\qquad=-\eta/B_0}
\\[5pt]
\displaystyle
\quad
 {}+B_0\bigg[\rho_1\varphi_{1\tau}-\rho_2\varphi_{2\tau} -\rho_1|\psi_{1x}^{(0)}|^2 +\rho_2|\psi_{2x}^{(0)}|^2
\\[10pt]
\displaystyle
\qquad
 {}+\rho_1\Big(A_1\psi_{1x}^{(1)\ast}+c.c.
 -A_1\frac{h_1^2}{2}\psi_{1xxx}^{(0)\ast}+c.c.\Big)
 \\[10pt]
\displaystyle
 \quad
 {}-\rho_2\Big(A_2\psi_{2x}^{(1)\ast}+c.c. -A_2\frac{h_2^2}{2}\psi_{2xxx}^{(0)\ast}+c.c.\Big)\bigg]=\eta_{xx}\,.
 \end{array}
\label{eqa21-1}
\end{equation}
Substituting $\psi_j^{(n)}$ from (\ref{eqa16})--(\ref{eqa19}) and using (\ref{eqa14}), one can rewrite the last equation as
\begin{gather}
 -\frac{B_1}{B_0}\eta+B_0\Bigg[\frac{h_1\rho_2+h_2\rho_1}{h_2}\varphi_{1\tau}
\nonumber\\[5pt]
\qquad {}
 -\frac{3\rho_1\rho_2(\rho_2-\rho_1)^3(h_1+h_2)^2}{4(h_1\rho_2+h_2\rho_1)^4}\eta^2
\label{eqa21-2}
\\[5pt]
\quad {}
 +\frac{\rho_1\rho_2(\rho_2-\rho_1)^2(h_1+h_2)^2(h_1^3\rho_2+h_2^3\rho_1)}{6(h_1\rho_2+h_2\rho_1)^4}\eta_{xx}\Bigg]=\eta_{xx}\,.
\nonumber
\end{gather}
Together with Eq.~(\ref{eqa15}) the latter equation form the final {\em system of governing equations for long-wavelength perturbations} of the flat-interface state:
\begin{equation}
\left\{
 \begin{array}{rcl}
 \displaystyle
 B_0\frac{h_1\rho_2+h_2\rho_1}{h_1h_2}(h_1\varphi_1)_{\tau}&
 \displaystyle =&
 \displaystyle \Bigg[1-\frac{1}{3}\frac{h_1^3\rho_2+h_2^3\rho_1}
 {h_1\rho_2+h_2\rho_1}\Bigg]\eta_{xx}
\\[15pt]
  &&\displaystyle
  {}+\frac{3}{2}\frac{\rho_2-\rho_1}{h_1\rho_2+h_2\rho_1}\eta^2
  +\frac{B_1}{B_0}\eta\,,
\\[15pt]
  \displaystyle
 \eta_{\tau}&
 \displaystyle =&
 \displaystyle -(h_1\varphi_1)_{xx}\,.
 \end{array}
\right.
\label{eqa22}
\end{equation}

Notice, equation system~(\ref{eqa22}) is valid for $B_1$ small compared to $B_0$, otherwise one cannot stay within the long-wa\-ve\-length approximation. It is only rarely possible to use long-wa\-ve\-length approximation for finite deviations from the linear instability threshold and derive certain information on the system dynamics \citep{Goldobin-Lyubimov-2007} or its properties \citep{Pimenova-etal-2015,Goldobin-etal-2015}; typically the long-wa\-ve\-length analysis is inconclusive for such conditions.

\subsection{Derivation of equations for 3-dimensional flow}
In this section we consider interfacial waves and flows which are non-uniform along the $y$-axis, the horizontal direction perpendicular to the vibration direction.

\subsubsection{The case of similar scales of patterns along the $x$- and $y$-directions ($\partial_y\sim\partial_x$)}
Firstly, let us consider the case of $\zeta=\zeta(x,y)$ and $\Phi_j=\Phi_j(x,y)$ for which $x$- and $y$- derivatives are of the same order of magnitude. After appropriate changes to equations, Eq.~(\ref{eqa17}) turns into
\begin{equation}
\Delta_2\psi^{(0)}=-\frac{\rho_1\rho_2(\rho_2-\rho_1)(h_1+h_2)}{2(h_1\rho_2+h_2\rho_1)^2}\eta_x\,,
\label{eqb17}
\end{equation}
where $\Delta_2$ stands for the Laplace operator with respect to the horizontal coordinates, $\Delta_2\equiv\partial_x^2+\partial_y^2$. Eq.~(\ref{eqa20-0}) remains unchanged. However, with Eq.~(\ref{eqb17}), after differentiation with respect to $x$, it yields a new solvability condition;
$$
-\frac{B_0\rho_1\rho_2(\rho_2-\rho_1)^2(h_1+h_2)^2}{2(h_1\rho_2+h_2\rho_1)^3}\psi_{xx}^{(0)}
 +\Delta_2\psi^{(0)}=0\,.
$$

The new solvability condition depends on the wave pattern. Specifically, for plane waves $\psi^{(0)}(x,y)\propto e^{i(k_x x+k_y y)}$ an analogue of Squire's theorem appears to be valid:
\begin{equation}
B_0(\beta)=\frac{B_0(0)}{\cos^2\beta}\,,
\label{eqb20}
\end{equation}
where $\beta$ is the angle between the wavevector ${\bf k}$ and the vibration direction, $B_0(0)$ is the linear instability threshold for 2-dimensional waves, which is determined by Eq.~(\ref{eqa20}). From Eq.~(\ref{eqb20}), the threshold of instability to 3-dimensional waves is increased for a finite value as compared to that for 2-d waves. Hence, 3-dimensional waves with similar scales along $x$- and $y$-directions can be considered, only when the system is already unstable with respect to 2-d waves. Since the latter instability leads to an explosive growth of practically any waves~\citep{Goldobin-etal-EPL-2014-solitons,Goldobin-etal-PRE-2015}, the study for 3-d waves under these circumstances is of marginal interest.

\subsubsection{The case of $\partial_y^2\sim\omega^{-1}\partial_x^2$}
According to Eq.~(\ref{eqb20}), the increase of the linear stability threshold for small $\beta$ is small and can be comparable to the values $B_1$ admissible within the framework of our theoretical analysis. The case of plane waves with small $\beta$ corresponds to wave patterns with $|\partial_y\eta|\ll|\partial_x\eta|$. For this case, one can consider the system dynamics beyond $(x,z)$-geometry and this dynamics will be still of physical significance (in contrast to the case of $\partial_y\sim\partial_x$), because it is observed below the threshold of the system instability. In what follows we consider the case of $\partial_y^2\sim\omega^{-1}\partial_x^2$.

Basic principles of the derivations for the 3-d case are similar to that for the case of $(x,z)$-geometry. Therefore, we do not provide a complete derivation for the 3-d case, but only discuss the equations which change compared to the 2-d case.

In Eq.~(\ref{eqa03}), additional non-negligible terms appear:
\begin{gather}
 B\Bigg[\frac{\rho_1a_1^2-\rho_2a_2^2}{2}+\rho_1\dot{\Phi}_1
 -\frac{\rho_1(h_1-\zeta)^2}{2}\dot{\Phi}_{1xx}
\nonumber\\[5pt]
\qquad
 {}-\frac{\rho_1}{2}\left(a_1-\Phi_{1x}+\frac{1}{2}(h_1-\zeta)^2\Phi_{1xxx}\right)^2
 -\frac{\rho_1}{2}\left(\Phi_{1y}\right)^2
\nonumber\\[5pt]
\qquad
 {}-\frac{\rho_1}{2}\left((h_1-\zeta)\Phi_{1xx}\right)^2
 -\rho_2\dot{\Phi}_2 +\frac{\rho_2(h_2+\zeta)^2}{2}\dot{\Phi}_{2xx}
\nonumber\\[5pt]
\qquad
 {}+\frac{\rho_2}{2}\left(a_2-\Phi_{2x}+\frac{1}{2}(h_2+\zeta)^2\Phi_{2xxx}\right)^2
  +\frac{\rho_2}{2}\left(\Phi_{2y}\right)^2
\nonumber\\[5pt]
\quad
 {}+\frac{\rho_2}{2}\left((h_2+\zeta)\Phi_{2xx}\right)^2
 +\dots\Bigg]+\zeta=\frac{\zeta_{xx}}{(1+\zeta_x^2)^{3/2}}\,.
\label{eqa03-3d}
\end{gather}
The kinematic conditions~(\ref{eqa05}) and (\ref{eqa06}) turn into
\begin{gather}
\dot\zeta=\left(-(h_1-\zeta)\Phi_{1x}+\frac{1}{3!}h_1^3\Phi_{1xxx} -a_1\zeta+\dots\right)_x
\nonumber\\[5pt]
\hspace{4cm}
  {}-h_1\Phi_{1yy}+\dots\,,
\label{eqa05-3d}
\end{gather}
\begin{gather}
\dot\zeta=\left((h_2+\zeta)\Phi_{2x}-\frac{1}{3!}h_2^3\Phi_{2xxx}  -a_2\zeta+\dots\right)_x
\nonumber\\[5pt]
\hspace{4cm}
  {}+h_2\Phi_{2yy}+\dots\,.
\label{eqa06-3d}
\end{gather}
Eqs.~(\ref{eqa03-3d})--(\ref{eqa06-3d}) form a self-contained equation system.

Collecting in Eqs.~(\ref{eqa05-3d}) and (\ref{eqa06-3d}) terms with $e^{i\omega t}$, one finds (in place of Eqs.~(\ref{eqa07}) and (\ref{eqa08}))
\begin{gather}
i\omega\xi+\xi_\tau=\Big(-(h_1-\eta)\psi_{1x}+\frac{1}{3!}h_1^3\psi_{1xxx}
\nonumber\\[5pt]
\hspace{2cm}
 {}
 +\xi\varphi_{1x}-A_1\eta+\dots\Big)_x -h_1\psi_{1yy}+\dots\,,
\label{eqa07-3d}
\end{gather}
\begin{gather}
i\omega\xi+\xi_\tau=\Big((h_2+\eta)\psi_{2x}-\frac{1}{3!}h_2^3\psi_{2xxx}
\nonumber\\[5pt]
\hspace{2cm}
 {}
 +\xi\varphi_{2x}-A_2\eta+\dots\Big)_x +h_2\psi_{2yy}+\dots\,.
\label{eqa08-3d}
\end{gather}
Constant with respect to $t$ terms sum-up to
\begin{gather}
\eta_\tau=\Big(-(h_1-\eta)\varphi_{1x}+\xi\psi_{1x}^\ast+c.c.-A_1\xi^\ast+c.c.+\dots\Big)_x
\nonumber\\[5pt]
\hspace{5cm}
 {}
 -h_1\varphi_{1yy}+\dots\,,
\label{eqa09-3d}
\end{gather}
\begin{gather}
\eta_\tau=\Big((h_2+\eta)\varphi_{2x}+\xi\psi_{2x}^\ast+c.c.-A_2\xi^\ast+c.c.+\dots\Big)_x
\nonumber\\[5pt]
\hspace{5cm}
 {}
 +h_2\varphi_{2yy}+\dots\,.
\label{eqa10-3d}
\end{gather}

The difference of (\ref{eqa08-3d}) and (\ref{eqa07-3d}) reads
$$
\begin{array}{r}
\displaystyle
\Big(h_1\psi_{1x}+h_2\psi_{2x}-\eta(\psi_1-\psi_2)_x
 -\frac{1}{6}(h_1^3\psi_1+h_2^3\psi_2)_{xxx}
\quad\\[10pt]
\displaystyle
 {}
 -\xi(\varphi_1-\varphi_2)_x +(A_1-A_2)\eta+\dots\Big)_x
\quad\qquad\\[10pt]
\displaystyle
 {}
 +\Big(h_1\psi_{1y}+h_2\psi_{2y}+\dots\Big)_y=0\,,
\end{array}
$$
which yields in the second orders of smallness, in place of Eq.~(\ref{eqa13}),
\begin{gather}
\left(h_1\psi_1^{(1)}+h_2\psi_2^{(1)}\right)_{xx}=\Big((\psi_1^{(0)}-\psi_2^{(0)})_x\eta
\nonumber\\[10pt]
 \;
 {}+\frac{1}{6}(h_1^3\psi_1^{(0)}+h_2^3\psi_2^{(0)})_{xxx}\Big)_x
-\Big(h_1\psi_1^{(0)}+h_2\psi_2^{(0)}\Big)_{yy}\,.
\label{eqa13-3d}
\end{gather}

Further, Eqs.~(\ref{eqa18}) and (\ref{eqa13-3d}) yield
\begin{eqnarray}
\psi_{1xx}^{(1)}&=&\bigg(\frac{\rho_2-\rho_1}{\rho_1(h_1\rho_2+h_2\rho_1)}\psi_x^{(0)}\eta
\nonumber\\[5pt]
&&
 {}
 +\frac{3h_2h_1^2\rho_1-2h_2^3\rho_1+h_1^3\rho_2}{6\rho_1(h_1\rho_2+h_2\rho_1)}\psi_{xxx}^{(0)}\bigg)_x
 -\frac{1}{\rho_1}\psi_{yy}^{(0)}
\nonumber\\[5pt]
 &=&\bigg(-\frac{\rho_2(\rho_2-\rho_1)^2(h_1+h_2)}{2(h_1\rho_2+h_2\rho_1)^3}\eta^2
\nonumber\\[5pt]
 &&\hspace{-1cm}
 {}-\frac{\rho_2(3h_2h_1^2\rho_1-2h_2^3\rho_1+h_1^3\rho_2)(\rho_2-\rho_1)(h_1+h_2)}
 {12\,(h_1\rho_2+h_2\rho_1)^3}\eta_{xx}\bigg)_x
\nonumber\\[5pt]
&&
 {}
 -\frac{1}{\rho_1}\psi_{yy}^{(0)}\,,
\label{eqa19-3d}
\end{eqnarray}
\begin{eqnarray}
\psi_{2xx}^{(1)}&=&\bigg(-\frac{\rho_1(\rho_2-\rho_1)^2(h_1+h_2)}{2(h_1\rho_2+h_2\rho_1)^3}\eta^2
\nonumber\\[5pt]
 &&\hspace{-1cm}
 {}-\frac{\rho_1(3h_1h_2^2\rho_2-2h_1^3\rho_2+h_2^3\rho_1)(\rho_2-\rho_1)(h_1+h_2)}
 {12\,(h_1\rho_2+h_2\rho_1)^3}\eta_{xx}\bigg)_x
\nonumber\\[5pt]
&&
 {}
 -\frac{1}{\rho_2}\psi_{yy}^{(0)}\,.
\label{eqa19-2-3d}
\end{eqnarray}

Eq.~(\ref{eqa21-1}) remains unchanged; however, one should substitute fields $\psi_j^{(1)}$ determined by Eqs.~(\ref{eqa19-3d}) and (\ref{eqa19-2-3d}) into it. Then, in place of Eq.~(\ref{eqa21-2}), one obtains
\begin{gather}
 -\frac{B_1}{B_0}\eta+B_0\left[\frac{h_1\rho_2+h_2\rho_1}{h_2}\varphi_{1\tau}
 \right.
\nonumber\\[10pt]
\qquad {}
 -\frac{3\rho_1\rho_2(\rho_2-\rho_1)^3(h_1+h_2)^2}{4(h_1\rho_2+h_2\rho_1)^4}\eta^2
\nonumber\\[10pt]
\qquad\qquad {} +\frac{\rho_1\rho_2(\rho_2-\rho_1)^2(h_1+h_2)^2(h_1^3\rho_2+h_2^3\rho_1)}{6(h_1\rho_2+h_2\rho_1)^4}\eta_{xx}
\nonumber\\[10pt]
\qquad\qquad\qquad \left.{}
 +\frac{1}{B_0}\partial_x^{-2}\eta_{yy}\right]=\eta_{xx}\,.
\label{eqa21-2-3d}
\end{gather}
In an infinite space with the condition that $\eta(x,y,\tau)$ remains finite, the action of operator $\partial_x^{-2}$ is well-defined.
Together with equation~(\ref{eqa15}), which remains unchanged, the latter equation form the final {\em system of governing equations for 3-d long-wavelength perturbations} of the flat-interface state:
\begin{equation}
\left\{
 \begin{array}{rcl}
 \displaystyle
 B_0\frac{\tilde{h}_1\tilde\rho_2+\tilde{h}_2\tilde\rho_1}{\tilde{h}_1\tilde{h}_2} \tilde{h}_1\tilde\varphi_{1\tilde\tau}&
 \displaystyle =&
 \displaystyle \left[1-\frac{1}{3}\frac{\tilde{h}_1^3\tilde\rho_2+\tilde{h}_2^3\tilde\rho_1}
 {\tilde{h}_1\tilde\rho_2+\tilde{h}_2\tilde\rho_1}\right]\tilde\eta_{\tilde x\tilde x}
\\[10pt]
 &&\displaystyle\hspace{-12mm}
 {}
 +\frac{3}{2}\frac{\tilde\rho_2-\tilde\rho_1}{\tilde{h}_1\tilde{\rho}_2+\tilde{h}_2\tilde{\rho}_1} \tilde\eta^2
 +\frac{B_1}{B_0}\tilde\eta-\partial_{\tilde{x}}^{-2}\tilde\eta_{\tilde{y}\tilde{y}}\,,
\\[15pt]
  \displaystyle
 \tilde\eta_{\tilde\tau}&
 \displaystyle =&
 \displaystyle -\tilde{h}_1\tilde\varphi_{1\tilde{x}\tilde{x}}\,.
 \end{array}
\right.
\label{eqa22-3d-dim}
\end{equation}
Here the dimensionless variables and parameters are marked with the tilde sign to distinguish them from original dimensional variables and parameters. (Above in this section, the tilde sign was omitted.) For convenience we explicitly specify how to read rescaling~(\ref{rescaling1}) with the tilde-notation:
 $x=L\tilde x$, $t=(L/b)\tilde t$, $\rho_i=\rho_\ast\tilde\rho_i$, etc.

\subsection{Conditions for applicability of the long-wavelength approximation}
On the basis of earlier works of ~\citet{Lyubimov-Cherepanov-1986} and the results for the case $h_1=h_2$~\citep{Goldobin-etal-PRE-2015}, we expected the linear instability of the flat-interface state to be long-wavelength for thin enough layers and relayed on this expectation. Now we can see an explicit quantitative condition for the layer to be ``thin enough''.
The exponential growth rate $\tilde\lambda$ of linear normal perturbations $(\tilde\eta,\tilde\varphi_1)\propto\exp[\tilde\lambda\tilde t+i(\tilde k_x\tilde x+\tilde k_y\tilde y)]$ of the trivial state obeys
\begin{equation}
\tilde\lambda^2=\frac{\tilde{h}_1\tilde{h}_2
\left(
 -\left[1-\frac{1}{3}\frac{\tilde{h}_1^3\tilde\rho_2+\tilde{h}_2^3\tilde\rho_1}
 {\tilde{h}_1\tilde\rho_2+\tilde{h}_2\tilde\rho_1}\right]\tilde{k}_x^4
 +\frac{B_1}{B_0}\tilde{k}_x^2-\tilde{k}_y^2\right)
}
 {B_0(\tilde{h}_1\tilde\rho_2+\tilde{h}_2\tilde\rho_1)}.
\label{eq-incr}
\end{equation}
From Eq.~(\ref{eq-incr}) one can see that the long-wa\-ve\-length perturbations are the most dangerous one and grow for $B_1>0$, if the expression in the square brackets is nonnegative,
\begin{equation}
\frac{\tilde{h}_1^3\tilde\rho_2+\tilde{h}_2^3\tilde\rho_1}
 {\tilde{h}_1\tilde\rho_2+\tilde{h}_2\tilde\rho_1}\le3\,.
\label{eq-long}
\end{equation}
Otherwise, the long-wa\-ve\-length perturbations are not the most dangerous one. Eq.~(\ref{eq-long}) imposes limitation from above on $h_1$ and $h_2$. This analysis of equation system~(\ref{eqa22-3d-dim}) only highlights the long-wavelength character of the linear instability, since it deals with the limit of small $\tilde{k}$ and does not provide information on the linear stability for finite $\tilde{k}$. A comprehensive proof of the long-wavelength character of the instability comes from the research by~\citet{Lyubimov-Cherepanov-1986}.

\subsection{Final rescaling and equations}
For consideration of the system dynamics below the linear instability threshold, i.e.\ for negative $B_1$, it is convenient to make further rescaling of coordinates and variables:
\begin{equation}
\begin{array}{c}
\displaystyle
 \tilde x\to x\sqrt{\frac{B_0}{(-B_1)}\left[1-\frac{1}{3}\frac{\tilde{h}_1^3\tilde\rho_2+\tilde{h}_2^3\tilde\rho_1}
 {\tilde{h}_1\tilde\rho_2+\tilde{h}_2\tilde\rho_1}\right]}\,,
\\[15pt]
\displaystyle
 \tilde y\to y\frac{B_0}{(-B_1)}\sqrt{1-\frac{1}{3}\frac{\tilde{h}_1^3\tilde\rho_2+\tilde{h}_2^3\tilde\rho_1}
 {\tilde{h}_1\tilde\rho_2+\tilde{h}_2\tilde\rho_1}}\,,
\\[15pt]
\displaystyle
 \tilde t\to t\sqrt{\frac{\tilde{h}_1\tilde\rho_2+\tilde{h}_2\tilde\rho_1}{\tilde{h}_1\tilde{h}_2}
 \frac{B_0^3}{B_1^2}
 \left[1-\frac{1}{3}\frac{\tilde{h}_1^3\tilde\rho_2+\tilde{h}_2^3\tilde\rho_1}
 {\tilde{h}_1\tilde\rho_2+\tilde{h}_2\tilde\rho_1}\right]}\,,\\[15pt]
\displaystyle
 \tilde\eta\to\eta\frac{\tilde{h}_1\tilde\rho_2+\tilde{h}_2\tilde\rho_1}{\tilde\rho_2-\tilde\rho_1}
 \frac{(-B_1)}{B_0}\,,\\[15pt]
\displaystyle
 \tilde\varphi_1\to\frac{\varphi}{\tilde{h}_1}
 \frac{\tilde{h}_1\tilde\rho_2+\tilde{h}_2\tilde\rho_1}{\tilde\rho_2-\tilde\rho_1}
 \left[1-\frac{1}{3}\frac{\tilde{h}_1^3\tilde\rho_2+\tilde{h}_2^3\tilde\rho_1}
 {\tilde{h}_1\tilde\rho_2+\tilde{h}_2\tilde\rho_1}\right]\,.
 \end{array}
\label{eqd01}
\end{equation}
Notice, that this means the following transformation of {\it initial dimensional} coordinates and variables:
\begin{equation}
\begin{array}{c}
\displaystyle
 x\to x\,L\sqrt{\frac{B_0}{(-B_1)}\left[1-\frac{1}{3L^2}\frac{h_1^3\rho_2+h_2^3\rho_1}
 {h_1\rho_2+h_2\rho_1}\right]}\,,
\\[15pt]
\displaystyle
 y\to y\,L\frac{B_0}{(-B_1)}\sqrt{1-\frac{1}{3L^2}\frac{h_1^3\rho_2+h_2^3\rho_1}
 {h_1\rho_2+h_2\rho_1}}\,,
\\[15pt]
\displaystyle
 t\to t\sqrt{\frac{h_1\rho_2+h_2\rho_1}{\rho_\ast h_1h_2}
 \frac{L^3 B_0^3}{b^2B_1^2}
 \left[1-\frac{1}{3L^2}\frac{h_1^3\rho_2+h_2^3\rho_1}
 {h_1\rho_2+h_2\rho_1}\right]}\,,\\[15pt]
\displaystyle
 \eta\to\eta\frac{h_1\rho_2+h_2\rho_1}{\rho_2-\rho_1}
 \frac{(-B_1)}{B_0}\,,\\[15pt]
\displaystyle
 \varphi_1\to\varphi\frac{bL}{h_1}
 \frac{h_1\rho_2+h_2\rho_1}{\rho_2-\rho_1}
 \left[1-\frac{1}{3L^2}\frac{h_1^3\rho_2+h_2^3\rho_1}
 {h_1\rho_2+h_2\rho_1}\right]\,.
 \end{array}
\label{eqd02}
\end{equation}

After the rescaling, equation system~(\ref{eqa22-3d-dim}) ({\it below the linear instability threshold}) takes a parameterless form;
\begin{equation}
\ddot\eta-\eta_{xx}-\eta_{yy} +\left(\frac{3}{2}\eta^2+\eta_{xx}\right)_{xx}=0\,.
\label{eqd03}
\end{equation}

The derivation of Eq.~(\ref{eqd03}) itself is one of the principle results we report with this paper, as it allows considering the 3-d evolution of quasi-steady patterns in the two-layer liquid system under the action of the vibration field for arbitrary ratio $h_1/h_2$ (in~\citep{Goldobin-etal-PRE-2015} the governing equation was derived for 2-d flows and $h_1=h_2$).

\section{Dynamics above the threshold and for inverted state}
Equation system~(\ref{eqa22-3d-dim}) derived in the previous section is relevant in the vicinity of the linear instability threshold, both below and above the threshold. Above the threshold, $B_1$ is positive and one has to replace $(-B_1)$ with $B_1$ in rescalings (\ref{eqd01})--(\ref{eqd02}), which leads to the change of the sign of the second term in Eq.~(\ref{eqd03}). Thus, {\it above the linear instability threshold}, equation system~(\ref{eqa22-3d-dim}) yields
\begin{equation}
\ddot\eta+\eta_{xx}-\eta_{yy} +\left(\frac{3}{2}\eta^2+\eta_{xx}\right)_{xx}=0\,.
\label{eq11}
\end{equation}

Furthermore, the derivation procedure can be formally repeated for the case where the heavy liquid overlies the light one (we refer to this system state as `inverted state'). The correction of the results for this case is straightforward; one can substitute $g\to-g$ in all the equations of the previous section, but make proper account for rescaling (\ref{rescaling1}), where the real-valued length scale $L=\sqrt{\alpha/[(\rho_2-\rho_1)g]}$ is to be kept unchanged. After this correction, one finds the same value of $B_0$ for the critical point of the linear stability of the flat interface state (generally, this does not necessarily mean that the state is stable on one of the sides of the point: it can be unstable on the both sides, but with different instability mechanisms). Within the vicinity of this point, one obtains dimensionless governing equations:
\\
$\bullet$~{\it below} the critical point $B_0$ for the {\it inverted state},
\begin{equation}
\ddot\eta+\eta_{xx}+\eta_{yy} -\left(\frac{3}{2}\eta^2+\eta_{xx}\right)_{xx}=0\,;
\label{eq12}
\end{equation}
\\
$\bullet$~{\it above} the critical point $B_0$ for the {\it inverted state},
\begin{equation}
\ddot\eta-\eta_{xx}+\eta_{yy} -\left(\frac{3}{2}\eta^2+\eta_{xx}\right)_{xx}=0\,.
\label{eq13}
\end{equation}

For understanding of the general properties of the system dynamics, it is interesting to compare Eqs.~(\ref{eqd03})--(\ref{eq13}) with the Boussinesq equation for gravity waves in shallow water.

\subsection{Two-dimensional flow ($\partial_y=0$)}
It was demonstrated above that the physical-space scale of variation of fields in the $y$-direction is large compared to that in the $x$-direction; therefore, a large lateral horizontal size of container is required to have room for variation of fields with $y$. Hence, when the lateral size of the container is non-large, the consideration can be restricted to the case of a 2-d flow. For $\partial_y=0$, Eqs.~(\ref{eqd03})--(\ref{eq13}) read
\begin{equation}
\ddot\eta+s_bs_g\eta_{xx}+s_g\left(\frac{3}{2}\eta^2+\eta_{xx}\right)_{xx}=0\,,
\label{eq14}
\end{equation}
where $s_b=+1$ and $-1$ above and below $B_0$, and $s_g=+1$ and $-1$ for the `normal' and inverted states, respectively.

Let us compare Eq.~(\ref{eq14}) with the original Boussinesq equation (OBE) for gravity waves in shallow water in the dimensional form (without vibrations and surface tension) \citep{Boussinesq-1872};
\begin{equation}
\ddot\eta-gh\eta_{xx}-gh\left(\frac{3}{2}\frac{\eta^2}{h}+\frac{h^2}{3}\eta_{xx}\right)_{xx}=0\,,
\label{eq15}
\end{equation}
where $h$ is the layer thickness. In the dimensionless form, the same equation is relevant for a two-layer system without vibrations~\citep{Choi-Camassa-1999}.

The interface dynamics for the normal state below $B_0$ ($s_b=-1$ and $s_g=+1$) corresponds to the `plus' Boussinesq equation~\citep{Manoranjan-etal-1988,Bogdanov-Zakharov-2002}, which differs from (\ref{eq15}), and was thoroughly discussed by \citet{Goldobin-etal-EPL-2014-solitons,Goldobin-etal-PRE-2015}. Noticeably, the dynamics above $B_0$ ($s_b=+1$ and $s_g=+1$) corresponds to Eq.~(\ref{eq15}) with negative $g$, which describes the falling of a liquid layer covering a ceiling; the flat-interface state is obviously unstable without any saturation for perturbation growth.

One can notice that the interface dynamics for the inverted state above $B_0$ is governed by OBE. However, this does not mean that the flat-interface state becomes stable as it is for a shallow water layer subject solely to the gravity field. Indeed, Eq.~(\ref{eq14}) is rigorously derived for the vibrational system under consideration and the short-wa\-ve\-length instability inherent to OBE (it can be seen also from Eq.~(\ref{eq-incr})) represents the properties of the real physical system. Meanwhile, the validity of OBE is restricted to the description of the gravity waves which are of long wa\-ve\-length not only for the original fluid dynamics equations, but also for Boussinesq equation~(\ref{eq14}).

Let us elaborate on the latter statement. Equation system~(25) in~\citep{Boussinesq-1872} for gravity wa\-ves in shallow water reads in our terms as
\begin{equation}
\left\{
\begin{array}{rcl}
\displaystyle\dot{\eta}+\varphi_{xx}&\!\!=\!\!&
 \displaystyle-(\eta\,\varphi_x)_x+\frac{1}{6}\varphi_{xxxx}\,,
\\[10pt]
\displaystyle\dot{\varphi}+\eta&\!\!=\!\!&
 \displaystyle-\frac{1}{2}(\varphi_x)^2+\frac{1}{2}\dot\varphi_{xx}\,,
\end{array}
\right.
\label{eq16}
\end{equation}
where the terms in the r.h.s.\ of equations are small, i.e., both nonlinearity and dispersion are small. Dynamic system~(\ref{eq16}) does not posses a short-wa\-ve\-length instability. To the leading corrections owned by nonlinearity and dispersion, for waves traveling in one direction, the latter equation system can be recast as OBE. Thus, the short-wa\-ve\-length instability of OBE is not an inherent property of the long-wavelength gravity waves in physical systems, but the result of a rough approximation, while for the inverted state of the vibrational system above $B_0$ this instability actually exists.

To summarize, the normal flat-interface state below the threshold $B_0$ is free of instabilities to infinitesimal perturbations. For other three cases, the flat-interface state becomes linearly unstable. The similarity between equations for the inverted state above $B_0$ and the original Boussinesq equation for gravity waves does not mean the former system is as stable as the latter one. Nonetheless, this similarity is remarkable and indicates that the long-wa\-ve\-length dynamics of the inverted system subject to strong vibrations becomes as stable as the shallow liquid layer subject solely to gravity.

\subsection{Three-dimensional flow}
For the system of a large lateral extent in the physical space, flows can be inhomogeneous in the $y$-direction. For the case of vibrationless system, the $x$- and $y$-directions are equivalent, while for the vibrational system we consider the $y$-direction can be discriminated. Waves propagate in the $y$-direction without dispersion and nonlinearity (see Eqs.~(\ref{eqd03})--(\ref{eq13}); this is valid for the leading order of our expansion, where the propagation of waves in the $x$-direction is already essentially determined by both dispersion and nonlinearity). Moreover, the system dynamics for the normal state above the threshold $B_0$ does not match anymore the dynamics of the liquid layer falling from a ceiling; even roughly, the term $\eta_{yy}$ appears with negative sign in Eq.~(\ref{eq11}) and with positive sign in the 3-d version of Eq.~(\ref{eq15}) with negative $g$. Furthermore, Eq.~(\ref{eq13}) governing the inverted system above $B_0$ possesses terms $-\eta_{xx}$ and $+\eta_{yy}$, while for the gravity waves in shallow water these terms will be both with sign ``$-$''. Thus, all the similarities discussed for 2-d flows are broken in the 3-d case.


\section{Conclusion}
\label{sec_concl}
We have generalized the derivation of the governing equations of conservative dynamics of interfacial waves in a two-layer system of inviscid liquids subject to horizontal vibrations for the case of $h_1\ne h_2$. Even though the governing equations in the dimensionless form are the same as for $h_1=h_2$ \citep{Goldobin-etal-PRE-2015}, the procedure of the construction of expansion has revealed the appearance of a new group of contributions, which vanish for $h_1=h_2$ (see Eq.~(\ref{eqa18})). The equations are relevant for thin enough layers, $(h_1^3\rho_2+h_2^3\rho_1)/(h_1\rho_2+h_2\rho_1)<3\alpha/[(\rho_2-\rho_1)g]$, where the linear instability is long-wavelength.

Further, the consideration has been extended to the case of three-dimensional flows varying in the second horizontal direction. An analogue of Squire's theorem can be formulated here (Eq.~(\ref{eqb20})). According to this theorem, the instability threshold for plane perturbations increases with the angle between the wavevector and the vibration direction. Hence, close to the instability threshold of the most dangerous perturbations, only patters with a ``slow'' dependence on $y$ are of interest. The governing equations for these patterns have been derived as well; Eqs.~(\ref{eqd03})--(\ref{eq13}) describe the system dynamics below and above the threshold $B_0$ for both the normal and inverted states.

For a system confined in the $y$-direction, where the flows are two-dimensional, equations of the system dynamics can be compared to the Boussinesq equations for the gravity waves in shallow water. Noticeably, above the threshold, the system dynamics for the normal state is similar to the dynamics of the falling of a liquid layer from a ceiling, while the dynamics for the inverted state is reminiscent of the usual gravity waves. However, the latter similarity does not mean the vibrational system to become as stable as a horizontal liquid layer subject to the gravity field. The vibrational system possesses a short-wavelength instability, which is actually present in the system, while for the gravity waves in shallow water this instability is due to the expansion truncation and is observed beyond the domain of the applicability of the Boussinesq equation to the specific physical system. Nonetheless, the long-wavelength modes of instability of the inverted state become stabilized by vibrations.

For three-dimensional flows, in contrast to the case of the gravity waves in a vibration-free shallow water, two horizontal directions are not equivalent. The sign of $\eta_{yy}$-term in equations breaks the similarities in dynamics of some configurations of the vibrational systems and gravity waves in a vibration-free shallow water layer, which were observed for the 2-d case.

The work has been supported by the Russian Science Foundation (grant no.~14-21-00090).

\end{document}